\documentclass[12pt]{article}
\usepackage{natbib}
\newtheorem{theorem}{Theorem}
\begin{document}

\title{Singularity Theorems in General Relativity: Achievements and 
Open Questions.}
\author{Jos\'e M. M. Senovilla\\
\\
F\'{\i}sica te\'orica, Universidad del Pa\'{\i}s Vasco, Apartado 644, 
\\
48080 Bilbao, Spain, {\tt josemm.senovilla@ehu.es}}
\date{}
\maketitle
\section{Introduction}
In this short note, written by a theoretical physicist, not a 
historian, I would like to present a brief overview of the acclaimed 
singularity theorems, which are often quoted as one of the greatest 
theoretical accomplishments in General Relativity and Mathematical 
Physics. 

Arguably, the singularity theorems are one of the few, if not the 
only one, consequences of Einstein's greatest theory which was not 
foreseen, probably not even suspected, by its founder. Many other 
consequences came also to a scientific firm basis only after 
Einstein's demise; to mention a few outstanding examples, let me cite 
gravitational lensing, gravitational radiation in binary systems (as 
in the PSR B1913+16 system), or the variation of frequency in 
electromagnetic radiation travelling in a gravity field. {\em All of 
them}, however, had been explicitly predicted in a way or another by 
Einstein. 

On the contrary, the singularity theorems such as we understand them 
now (the result in \citep{E} is concerned with quite another type of 
singularities), the global developments needed for them, and the 
derived inferences, were not mentioned, neither directly nor 
indirectly, in any of his writings. This was so despite the various 
clear indications that the appearance of some kind of ``singularity", 
or catastrophical behaviour, was a serious possibility within the 
orthodox theory. For instance, the Friedman-Lema\^{\i}tre models 
\citep{Fri,Fri2,Lem}, generally accepted by the 1930's as providing 
explanation for the observed galactic (or nebulosae) redshifts, 
contain the famous ``creation  time" (Friedman's wording) identified 
by Friedman and Lema\^{\i}tre themselves, at which the space-time is 
simply destroyed. Or even grosser, the proof by Oppenheimer and 
Snyder \citep{OS} that the ``Schwarzschild surface"---which, as we 
know, is {\em not} a singularity, but a horizon, see for instance the 
well-documented historical review on this subject presented in 
\citep{TCE}--- was reachable and crossable by innocuous models 
containing realistic matter such as dust. (By the way, this dust 
eventually ends in a catastrophic (real) future singularity.) Not to 
mention the impressive result found by Chandrasekhar ---to which 
Eddington opposed furiously--- of the upper mass limit for stars in 
equilibrium, even when taking into account the quantum effects, 
implying that stars with a larger mass will inevitably collapse 
\citep{Ch}.

In spite of all these achievements, as I was saying, Einstein and the 
orthodoxy simply dismissed the catastrophic behaviours 
(singularities) as either a mathematical artifact due to the presence 
of (impossible) exact spherical symmetry, or as utterly impossible 
effects, scientifically untenable, obviously unattainable, beyond the 
feasibility of the physical world ---see e.g.\ \citep{E2}. Of course, 
this probably is certain in a deep sense, as infinite values of 
physical observables must not be accepted and every sensible 
scientist would defend similar sentences. {\em However}, one must be 
prepared to probe the limits of any particular theory, and this was 
simply not done with General Relativity at the time. It was necessary 
to wait for a new generation of physicists and mathematicians, 
without the old prejudices and less inhibited, and probably also less 
busy with the quantum revolution, who could finally take the question 
of the singularities, the gravitational collapse, and the past of the 
Universe, seriously within the theory.

Thus, one can say that the singularity theorems are the most genuine 
post-Einsteinian content of General Relativity.

\subsection{The Raychaudhuri equation}
Curiosuly enough, the first result concerning prediction of 
singularities under reasonable physical conditions, due to 
Raychaudhuri, came to light exactly the same year of Einstein's 
decease. In 1955 Raychaudhuri published what is considered the first 
singularity theorem, and included a version of the equation named 
after him which is the basis of later developments and of
{\em all} the singularity theorems \citep{R}. The Raychaudhuri 
equation can be easily derived (see for instance \citep{D} in a 
recent tribute to Raychaudhuri), as it has a 
very simple geometrical interpretation: take the well-known Ricci 
identity
\begin{equation}
(\nabla_{\mu}\nabla_{\nu}-\nabla_{\nu}\nabla_{\mu})u^{\alpha}=
R^{\alpha}{}_{\rho\mu\nu}u^{\rho}
\label{ric}
\end{equation}
which is mathematically equivalent to the standard definition of the 
Riemann tensor
$$
\left(\nabla_{\vec X}\nabla_{\vec Y}-\nabla_{\vec Y}\nabla_{\vec 
X}-\nabla_{[\vec X,\vec Y]}\right)\vec Z = R(\vec X, \vec Y)\vec Z 
\hspace{1cm} \forall \vec X, \vec Y, \vec Z
$$
and contract $\alpha$ with $\mu$ in (\ref{ric}), then with $u^{\nu}$, 
to obtain
$$
u^{\nu}\nabla_{\mu}\nabla_{\nu}u^{\mu}-u^{\nu}\nabla_{\nu}\nabla_{\mu}u^{\mu}=
R_{\rho\nu}u^{\rho}u^{\nu}
$$
where $R_{\rho\nu}$ is the Ricci tensor. Reorganizing by parts the 
first summand on the lefthand side one derives
\begin{equation}
u^{\nu}\nabla_{\nu}\nabla_{\mu}u^{\mu}+\nabla_{\mu}u_{\nu}\nabla^{\nu}u^{\mu}-
\nabla_{\mu}(u^{\nu}\nabla_{\nu}u^{\mu})+R_{\rho\nu}u^{\rho}u^{\nu}=0
\end{equation}
which is the Raychaudhuri equation. Raychaudhuri's important 
contribution amounts to understanding the physical implications of 
this relation. Observe that, in the case that $u^{\mu}$ defines a 
(affinely parametrized) {\em geodesic} vector field, then 
$u^{\nu}\nabla_{\nu}u^{\mu}=0$ and the third term vanishes. The 
second term can be rewritten by splitting 
$$
\nabla_{\mu}u_{\nu}=S_{\mu\nu}+A_{\mu\nu}
$$
into its symmetric $S_{\mu\nu}$ and antisymmetric $A_{\mu\nu}$ parts, 
so that
$$
\nabla_{\mu}u_{\nu}\nabla^{\nu}u^{\mu}=S_{\mu\nu}S^{\mu\nu}-A_{\mu\nu}A^{\mu\nu}\, 
.
$$
Now the point is to realize that (i) if $u^{\mu}$ is time-like (and 
normalized) or null, then both $S_{\mu\nu}S^{\mu\nu}$ and 
$A_{\mu\nu}A^{\mu\nu}$ are non-negative; and (ii) $u_{\mu}$ is 
proportional to a gradient (therefore defining orthogonal 
hypersurfaces) if and only if $A_{\mu\nu}=0$. In summary, for 
hypersurface-orthogonal geodesic time-like or null vector fields 
$u^{\mu}$ one has
$$
u^{\nu}\nabla_{\nu}\nabla_{\mu}u^{\mu}=
-S_{\mu\nu}S^{\mu\nu}-R_{\rho\nu}u^{\rho}u^{\nu}
$$
so that the sign of the derivative of the divergence 
$\nabla_{\mu}u^{\mu}$ along these geodesics is governed by the sign 
of $R_{\rho\nu}u^{\rho}u^{\nu}$. If the latter is non-negative, then 
the former is non-positive. In particular, if the divergence is 
negative at some point and $R_{\rho\nu}u^{\rho}u^{\nu}\geq 0$ then 
necessarily the divergence will reach an infinite negative value in 
finite affine parameter (unless all the quantities are zero 
everywhere).

If there are physical particles moving along these geodesics, then 
clearly a physical singularity is obtained, as the average volume 
decreases and the density of particles will be unbounded. This was 
the situation treated by Raychaudhuri for the case of irrotational 
dust. In general, no singularity is predicted, though, and one only 
gets a typical {\em caustic} along the flow lines of the congruence 
defined by $u^{\mu}$. This generic property is usually called the 
{\em focusing effect} on causal geodesics. For this to take place, of 
course, one needs the condition 
\begin{equation}
R_{\rho\nu}u^{\rho}u^{\nu}\geq 0 \label{sec}
\end{equation}
which is a {\em geometric} condition and independent of the 
particular theory. However, in General Relativity, one can relate the 
Ricci tensor to the energy-momentum tensor $T_{\mu\nu}$ via 
Einstein's field equations
\begin{equation}
R_{\mu\nu}-\frac{1}{2}g_{\mu\nu}R+\Lambda g_{\mu\nu}=\frac{8\pi 
G}{c^4}T_{\mu\nu}
\label{efe}
\end{equation}
where $R$ is the scalar curvature, $G$ is Newton's gravitational 
constant, $c$ is the speed of light in vacuum and $\Lambda$ the 
cosmological constant. Thereby, the condition (\ref{sec}) can be 
rewritten in terms of physical quantities. This is why sometimes 
(\ref{sec}), when valid for all time-like $u^{\mu}$, is called the 
{\em strong energy condition} \citep{HE}. One should bear in mind, 
however, that this is a condition on the Ricci tensor (a geometrical 
object) and therefore it will not always hold: see the discussion in 
Section \ref{appraisal} below and \citep[sect.~6.2]{S}.

An important remark of historical importance is that, before 1955, 
G\"odel wrote his famous paper \citep{G} in a volume dedicated to 
Einstein's 70th anniversary. This paper is considered 
\citep[see][sect.~3]{TCE} the genesis of many of the necessary 
techniques and some of the global ideas which were used in the path 
to the singularity theorems, specially concerning causality theory. 
For further information the reader is referred to \citep{Ell}. 
However, the subject was not ripe and had to wait, first, to the 
contribution by Raychaudhuri, and then, to the imaginative and 
fruitful ideas put forward by Roger Penrose in the 1960s.

\section{Remarks on singularities and extensions}
The problem of the definition of the concept of singularity in 
General Relativity is very difficult indeed, as can be appreciated by 
reading on its historical development 
\citep{HE,TCE}. The intuitive ideas are clear: if any physical or 
geometrical quantity blows up, this signals a singularity. However, 
there are problems of two kinds: 
\begin{itemize}
\item the singular points, by definition, do not belong to the 
space-time which is only constituted by regular points. Therefore, one 
cannot say, in principle, ``when'' or ``where'' is the singularity.
\item characterizing the singularities is also difficult, because the 
divergences (say) of the curvature tensor can depend on a bad choice 
of basis, and even if one uses only curvature invariants, independent 
of the bases, it can happen that all of them vanish and still there 
are singularities.
\end{itemize}
The second point is a genuine property of Lorentzian geometry, that 
is, of the existence of one axis of time of a different nature to the 
space axes. 

Therefore, the only sensible definition with a certain consensus 
within the relativity community is by ``signalling'' the 
singularities, ``pointing at them'' by means of quantities belonging 
to the space-time exclusively. And the best and simplest pointers are 
curves, of course. One can imagine what happens if a brave traveller 
approaches a singularity: he/she will disappear from our world in a 
{\em finite} time. The same, but time-reversal, can be imagined for 
the ``creation'' of the Universe: things {\em suddenly} appeared from 
nowhere a {\em finite} time ago. All in all, it seems reasonable to 
diagnose the existence of singularities whenever there are particles 
(be them real or hypothetical) which go to, or respectively come 
from, them and disappear unexpectedly or, respectively, subito come 
to existence.

And this is the basic definition of singularity \citep{Ger,HE}, the 
existence of {\em incomplete and inextensible} curves. That is to 
say, curves which cannot be extended in a regular manner within the 
space-time and do not take all possible values of their canonical 
parameter. Usually, only causal (time-like or null) curves are used, 
but in principle also incomplete space-like curves define 
singularities. The curves do not need to be geodesic, and as a matter 
of fact there are known examples of geodesically complete space-times 
with incomplete time-like curves of everywhere bounded acceleration 
\citep{Ger}. It must be remarked, however, that all singularity 
theorems prove merely the existence of {\em geodesic} incompleteness, 
which of course is a {\em sufficient} condition for the existence of 
singularities according to the definition.

Some fundamental questions, which sometimes are omitted, arise at 
this point. How can one give structure and properties to the 
singularities? What is the relation between geodesic incompleteness 
and curvature problems, if any? Singularities in the above sense 
clearly reach, or come from, the {\em edge} of space-time. This is 
some kind of boundary, or margin, which is not part of the space-time 
but that, somehow, it is accessible from within it. Thus the 
necessity of a rigurous definition of the boundary of a space-time. 
Given that a Lorentzian manifold is not a metric space in the usual 
sense, one cannot use the traditional completion procedure by means 
of Cauchy sequences. The most popular approach in this sense has been 
the attempt to attach a {\em causal boundary} to the space-time, see 
\citep{GS} for an up-to-date review. But this is not exempt of 
recurrent problems which have not been completely solved yet. 

Furthermore, the existence of incomplete geodesics which are not 
extensible in a given space-time may indicate {\em not} a problem with 
the curve and the geometrical properties along it when approaching 
the edge, but rather {\em incompleteness} of the space-time itself. 
For instance, flat space-time without the origin has infinite 
inextensible incomplete curves. These cases, however, are to be 
avoided, as one can easily see that the problem arises due to the 
excision of regular points. This is why one usually restricts 
considerations to inextensible space-times when dealing with 
singularity theorems \citep{HE}. The physical problem, however, is 
hidden under the carpet with this attitude because: what are we 
supposed to do with given extensible space-times? The answer may seem 
simple and easy: just extend it until you cannot extend it anymore. 
However, this is not so simple for several reasons 
\citep[see][sects.~3 and 7]{S}:
\begin{itemize}
\item extensions are not obvious, and generally not unique. Usually 
there are infinite inequivalent choices.
\item not even analytical extensions are unique in general, let alone 
smooth extensions.
\item it can happen that there are incomplete curves, no curvature 
problems, and no possible (regular) extension.
\item sometimes, for a given fixed extensible space-time, there are 
extensions leading to new extensible space-times, other extensions 
leading to singular space-times, and even other extensions which are 
free of singularities and inextensible. It may seem obvious that the 
last choice would be the preferred one by relativists, but this is 
simply not the case ---if the singularity-free extension violates a 
physical condition, such as causality or energy positivity, then the 
other extensions will be chosen---.
\item which physical reasons are to be used to discriminate between 
inequivalent extensions?
\end{itemize}
As a drastic example of the above, take the traditional case of the 
Schwarzschild solution which, as is known, is extensible through the 
horizon. In textbooks one usually encounters the {\em unique} maximal 
analytical vacuum extension due to Kruskal and Szekeres keeping 
spherical symmetry. However, if one drops any one of the conditions 
(vacuum, analyticity) many other maximal extensions are possible, see 
e.g. \citep{S} where at least eleven inequivalent extensions were 
explicitly given. This should make plain that the question of 
singularities is intimately related to the problem of how, why and to 
where a given extensible space-time must be extended.

\section{Singularity theorems: Critical appraisal}
\label{appraisal}
The first singularity theorem in the modern sense is due to Penrose 
\citep{P}, who in this seminal paper introduced the fundamental 
concept of {\em closed trapped surface} and opened up a new avenue of 
fruitful research. His main, certainly innovative idea, was to prove 
null geodesic incompleteness if certain {\em initial} conditions, 
reasonable for known physical states of collapse, were given {\em 
irrespective of any symmetry or similar restrictive properties}. 

Since then, many singularity theorems have been proven 
\citep[see][]{HE,S}, some of them applicable to cosmological 
situations, some to star or galaxy collapse, and others to the 
collision of gravitational waves. The culmination was the celebrated 
Hawking-Penrose  theorem \citep{HP}, which since then is the 
singularity theorem {\em par excellence}. However, all of the 
singularity theorems share a well-defined skeleton, the very same 
pattern. This is, succintly, as follows \citep{S}
\begin{theorem}[Pattern Singularity Theorem]
If a space-time of sufficient differentiability satisfies
\begin{enumerate}
\item a condition on the curvature
\item a causality condition
\item and an appropriate initial and/or boundary condition
\end{enumerate}
then there are null or time-like inextensible incomplete geodesics.
\end{theorem}

I have started by adding explicitly the condition of sufficient 
differentiability. This is often ignored, but it is important from 
both the mathematical and the physical points of view. A breakdown of 
the differentiability should not be seen as a true singularity, 
specially if the problem is mild and the geodesic curves are 
continuous. The theorems are valid if the space-time metric tensor 
field $g_{\mu\nu}$ is of class $C^2$ (twice differentiable with 
continuity), but they have not been proven in general if the first 
derivatives of $g_{\mu\nu}$ satisfy only the Lipshitz condition. This 
problem is of physical relevance, because the entire space-time of a 
star or a galaxy, say, is usually considered to have two different 
parts matched together at the surface of the body, and then the 
metric tensor is not $C^2$ at this surface: it cannot be, for there 
must exist a jump in the matter density which is directly related to 
the second derivatives of $g_{\mu\nu}$ via equations (\ref{efe}). As 
an example, the Oppenheimer-Snyder collapsing model does not satisfy 
the $C^2$ condition. A list of the very many places where this 
condition is used in the singularity theorems can be found in 
\citep[p.~799]{S}.

Then there is the ``curvature condition''. I have preferred this name 
rather than the usual {\em energy} and {\em generic} condition to 
stress the fact that this assumption is of a geometric nature, and it 
is absolutely indispensable: it enforces the geodesic focusing via 
the Raychaudhuri equation ---and other similar identities---. The 
majority of the theorems, specially the stronger ones, use the 
condition (\ref{sec}), which is usually called {\em strong energy 
condition} if it is valid for all time-like $u^{\mu}$, and {\em null 
convergence condition} if it is valid only for null vectors. The 
former name is due to the equivalent relation, via the field 
equations (\ref{efe})
$$
T_{\rho\nu}u^{\rho}u^{\nu}\geq \frac{c^4\Lambda}{8\pi 
G}-\frac{1}{2}T^{\rho}{}_{\rho}
$$
for unit time-like $u^{\mu}$. This involves energy-matter variables. 
However, this condition does {\em not} have to be satisfied in 
general by realistic physical fields. To start with, it depends on 
the sign of $\Lambda$. But furthermore, even if $\Lambda = 0$, the 
previous condition does not hold in many physical systems, such as 
for instance scalar fields \citep[p.~95]{HE}. As a matter of fact, 
most of the inflationary cosmological models violate the above 
condition. Let us stress that the physically compelling energy 
condition is the {\em dominant energy condition} \citep{HE}, but this 
has in principle nothing to do with the assumptions of the 
singularity theorems. In particular, there are many examples of 
reasonable singularity-free space-times satisfying the dominant, but 
not the strong, energy condition \citep[see e.g.][sect.7.3]{S}. In my 
opinion, this is one of the weak points of the singularity theorems. 

The ``causality condition'' is probably the most reasonable and 
well-founded, and perhaps also the less restrictive, condition. There 
are several examples of singularity theorems without any causality 
condition \citep[][Theorem 4, p.~272]{HE} \citep{MI}.
The causality condition is assumed for two types of reasons. Firstly, 
to prevent the possibility of avoiding the future, that is to provide 
a well-defined global time-arrow. This may seem superfluous, but it 
is known since the results in \citep{G} that there may be closed 
time-like lines, that is, curves for which the time passes to the 
future permanently and nevertheless reach their own past. And secondly, 
to ensure the existence of geodesics of maximal proper time between 
two events, and therefore geodesics without focal points 
\citep[][Prop.~4.5.8]{HE}.

Recapitulating, the first two conditions in the theorems imply
\begin{itemize}
\item the focusing of {\em all} geodesics, ergo the existence of 
caustics and focal points, due to the curvature condition
\item the existence of geodesics of maximal proper time, and 
therefore necessarily {\em without focal points}, joining any two 
events of the space-time
\end{itemize}
Obviously, a contradiction starts to glimmer if all geodesics are 
complete. There is not such yet, though, and this is because at this 
stage there is no finite {\em upper bound} for the proper time of 
selected families of time-like geodesics (and analogously for 
null geodesics). 

To get the final contradiction one needs to add the third condition. 
And this is why the initial/boundary condition is absolutely 
essential in the theorems. There are several possibilities for this 
condition, among them (i) an instant of time at which the whole 
Universe is strictly expanding, (ii) closed universes so that they 
contain space-like compact slices; (iii) the existence of closed (that 
is, compact without boundary) trapped surfaces. 

This last concept, due to Penrose \citep{P}, has become the most 
popular and probably the most important one, specially due to its 
relevance an applicability in many other branches of Relativity. 
Trapped surfaces are 2-dimensional differentiable surfaces (such as 
a donut's skin or a sphere) with a mean curvature vector which is 
time-like everywhere, or in simpler words, such that the initial 
variation of area along {\em any future direction} is always 
decreasing (or always increasing). An example of a non-compact 
trapped surface in flat space-time (in Cartesian coordinates) is given 
by
$$
e^t=\cosh z , \,\,\, x=\mbox{const.}
$$
There cannot be, however, compact trapped surfaces in flat space-time. 
An example of a compact trapped surface is given for instance by any 
2-sphere $t=$const., $x^2+y^2+z^2=R^2=$const., in a Friedman model
$$
ds^2 =-dt^2+a^2(t)\left(dx^2+dy^2+dz^2\right)
$$
as long as $R > 1/|\dot a|$, which is always possible. 

Whether or not the initial or boundary condition is realistic, or 
satisfied by the actual physical systems, is debatable. We will 
probably find it very difficult to ascertain if the Universe is 
spatially finite, or if the {\em whole} Universe is strictly 
expanding now. There is a wide agreement, however, that it is at 
least feasible that in some situations the third condition in the 
theorems will hold. For example, given the extremely high degree of 
isotropy observed in the cosmic microwave background radiation, one 
can try to infer that the Universe resembles a Friedman-Lema\^{\i}tre 
model, thus containing trapped spheres as the one shown above. 
Nevertheless, there are several way outs to this scheme of reasoning, 
for example a cosmological constant, or deviations from the model at 
distances (or redshifts) of the order or higher than the visible 
horizon. For a discussion see \citep[][sects.~6.4 and 7.2]{S}.

Let us finally pass to the conclusion of the theorems. In most 
singularity theorems this is just the existence of at least one 
incomplete causal geodesic. This is very mild, as it can be a mere 
localized singularity. This leaves one door open (extensions) and 
furthermore it may be a very mild singularity. In addition, the 
theorems do not say anything, in general, about the situation and the 
character of the singularity. We cannot know whether it is in the 
future or past, or whether it will lead to a blow-up of the curvature 
or not.

In the next section I am going to present a very simple solution of 
Einstein's field equations which shows explicitly the need of the 
``small print'' in the theorems, and that sometimes their assumptions 
are more demanding than generally thought.

\section{An illustrative singularity-free space-time}
Nowadays there are many known singularity-free space-times, some of 
them are spatially inhomogeneous universes, others contain a 
cosmological constant, and there are a wide variety of other types 
\citep[see][and references therein]{S}. Perhaps the most famous 
singularity-free ``cosmological model'' was presented in \citep{S2}, 
because it had some impact on the scientific community \citep[see 
e.g.][]{M} and opened up the door for many of the rest. The impact 
was probably due to the general belief that such a space-time was 
forbidden by the singularity theorems. But, of course, this was 
simply a misunderstanding, and it only meant that we had thought that 
the singularity theorems were implying more than they were actually 
saying.

The space-time has cylindrical symmetry, and in typical cylindrical 
coordinates $\{t,\rho,\varphi,z\}$ its line-element takes the form
\begin{eqnarray*}
ds^2=\cosh^4(act)\cosh^2(3a\rho)(-c^2dt^2+d\rho^2)+\hspace{2cm}\\
\frac{1}{9a^2}\cosh^4(act)\cosh^{-2/3}(3a\rho)\sinh^2(3a\rho)d\varphi^2+
\cosh^{-2}(act)\cosh^{-2/3}(3a\rho)dz^2
\end{eqnarray*}
where $a>0$ is a constant. This is a solution of the field equations 
(\ref{efe}) for $\Lambda =0$ and an energy-momentum tensor of 
perfect-fluid type
$$
T_{\mu\nu}=\rho u_{\mu}u_{\nu}+p\, (g_{\mu\nu}+u_{\mu}u_{\nu})
$$
where $\rho$ is the energy density of the fluid given by
$$
\frac{8\pi G}{c^4}\varrho = 15a^2\cosh^{-4}(act)\cosh^{-4}(3a\rho)\, ,
$$
$p$ its isotropic pressure and
$$
u_{\mu}=\left(-c\cosh^2(act)\cosh(3a\rho),0,0,0\right)
$$
defines the unit velocity vector field of the fluid. Observe that 
$u^{\mu}$ is not geodesic (except at the axis). The fluid has a 
realistic barotropic equation of state
$$
p=\frac{1}{3}\rho \, .
$$
This is the canonical equation of state for radiation-dominated 
matter and is usually assumed at the early stages of the Universe. 
Note that the density and the pressure are regular everywhere, and 
one can in fact prove that the space-time is completely free of 
singularities and geodesically complete \citep{CFS}.

The space-time satisfies the stronger causality conditions (it is 
globally hyperbolic), and also all energy conditions (dominant, 
strict strong). The fluid divergence is given by
\begin{equation}
\nabla_{\mu}u^{\mu} = 3a\frac{\sinh(act)}{\cosh^3(act)\cosh(3a\rho)} 
\label{exp}
\end{equation}
so that this universe is contracting for half of its history ($t<0$) 
and expanding for the second half ($t>0$), having a rebound at $t=0$ 
which is driven by the spatial gradient of pressure. Observe that the 
whole universe is expanding (that is, $\nabla_{\mu}u^{\mu}>0$) {\em 
everywhere} if $t>0$, and recall that this was one of the possibilities we 
mentioned for the third condition in the singularity theorems: an 
instant of time with a strictly expanding whole universe. Thus, how 
can this model be geodesically complete and singularity-free?

Well, the precise condition demanded by one of the theorems in 
globally hyperbolic space-times is that $\nabla_{\mu}u^{\mu}> b >0$ 
for a constant $b$. That is, $\nabla_{\mu}u^{\mu}$ has to be bounded 
from below by a positive constant. But this is not the case for 
(\ref{exp}), which is strictly positive everyhwere but not 
bounded from below by a positive constant because $\lim_{\rho \rightarrow 
\infty}\nabla_{\mu}u^{\mu}=0$. This minor, subtle, difference allows 
for the model to be singularity-free! All other possibilities for the 
initial/boundary condition in the several would-be applicable 
singularity theorems can be seen to {\em just} fail in the model, in 
a similar manner. For a complete discussion see \citep{CFS} and 
\citep[][sect.~7.6]{S}. One can also see that the focusing effect on 
geodesics takes place fully in this space-time, but nevertheless there 
is no problem with the existence of maximal geodesics between any 
pair of points \citep[see][pp.~829-830]{S}.

This simple model showed that there exist well-founded, well-behaved 
classical models expanding everywhere, satisfying all energy and 
causality conditions, and singularity-free. Of course, the model is 
not realistic in the sense that it cannot describe the actual 
Universe ---for instance, the isotropy of the cosmic background 
radiation cannot be explained---, but the question arises of whether 
or not there is room left over by the singularity theorems to 
construct geodesically complete {\em realistic} universes.

It should be stressed that this model does not describe a 
``cylindrical star'', because the pressure of the fluid does not 
vanish anywhere. Nevertheless, as can be seen from the previous 
formulae, for example the energy density is mainly concentrated in an 
area around the axis of symmetry, dying away from it very quickly as 
$\rho$ increases. This may somehow put some doubts about the 
relevance of this type of models for {\em cosmological} purposes. In 
this sense, there was an interesting contribution by Raychaudhuri 
himself \citep{R2}, where he tried to quantify this property in a 
mathematical condition. Unfortunately, this time he was not 
completely right \citep[see][]{S3}, because he used {\em space-time} 
averages of the physical variables such as the energy density. But 
one can easily seen that the vanishing of such averages is a property 
shared by the majority of the models, be them singular or not. 
However, his work provided some inspiration, and I believe that 
the vanishing of the {\em spatial} averages of the physical 
variables is probably a condition which may allow to distinguish 
between the singularity-free models allowed by the singularity 
theorems and the singular ones. A conjecture of the 
singularity-theorem type was put forward in \citep{S3} with this 
purpose. 

All in all, the main conclusion of this contribution is to remind 
ourselves that it is still worth to develop further, understand 
better, and study careful the singularity theorems, and their 
consequences for realistic physical systems.

\end{document}